\documentclass[12pt]{article}
\usepackage{epsf}
\def\simgt{\rlap{\lower 4 pt \hbox{$\mathchar \sim$}} \raise 1pt \hbox {$>$}}
\def\simlt{\rlap{\lower 4 pt \hbox{$\mathchar \sim$}} \raise 1pt \hbox {$<$}}

\begin{document}

\begin{titlepage}
\begin{flushright}
 CERN-TH/99-163\\
 \mbox{DTP/99/60\hspace*{0.45cm}}\\
 hep-ph/9906475\\ 
\end{flushright}
\vskip 1.8cm
\begin{center}
\boldmath
{\large\bf The bottom $\overline{\rm MS}$ quark mass from sum rules\\[0.2cm]
at next-to-next-to-leading order}
\unboldmath 
\vskip 1.4cm
{\sc M. Beneke}
\vskip .3cm
{\it Theory Division, CERN, CH-1211 Geneva 23, Switzerland}
\vskip 0.7cm
{\sc A. Signer}
\vskip .3cm
{\it Department of Physics, University of Durham, \\ 
Durham DH1 3LE, England}
\vskip 1cm

{\em (June, 22, 1999)}

\vskip 1cm
\end{center}

\begin{abstract}
\noindent 
We determine the bottom $\overline{\rm MS}$ quark mass 
$\overline{m}_b$ and the quark mass in the potential subtraction 
scheme from moments of the $b\bar{b}$ production cross section 
and from the mass of the Upsilon 1S state at next-to-next-to-leading 
order in a reorganized perturbative expansion that sums 
Coulomb exchange to all orders. We find 
$\overline{m}_b(\overline{m}_b)=(4.25\pm 0.08)\,$GeV and 
$m_{b,\rm PS}(2\,\mbox{GeV})=(4.59\pm 0.08)\,$GeV for the 
potential-subtracted mass at the scale $2\,$GeV, adopting a 
conservative error estimate.
\end{abstract}

\vfill

\end{titlepage}


{\em Introduction.}  
Accurate determinations of the bottom quark mass in perturbative QCD 
usually rely on properties of the spectrum of Upsilon mesons and 
$b\bar{b}$ production near threshold. Since already for the 
$\Upsilon(\mbox{1S})$ state the momentum scale 
$p=m_b C_F \alpha_s(p)/2 \sim 1.25\,$GeV and energy scale 
$E=m_b (C_F \alpha_s(p))^2/4 \,\,\simlt\,\, 0.3\,$GeV 
are too small for perturbation theory to be expected to to work, 
one considers 
averages over the $b\bar{b}$ production cross section through 
a virtual photon, including the $\Upsilon$ resonances\footnote{
The cross section is normalized such that $R_{b\bar{b}}(s)=N_c e_b^2=1/3$ 
in the ultra-relativistic limit, where $e_b$ is the bottom quark electric 
charge. $R_{b\bar{b}}(s)$ is related to the vector current 
two-point function $\Pi(q^2)$ in the usual way. 
The normalization factor on the left-hand side of 
(\ref{eq1}) is inserted for convenience.} \cite{NSVZ}:
\begin{equation}
\label{eq1}
M_n/(10\,\mbox{GeV})^{2 n}\equiv \frac{12\pi^2}{n!}\,
\frac{d^n}{d(q^2)^n}\,\Pi(q^2)_{\big| q^2=0} = 
\int\limits_0^\infty \!
\frac{ds}{s^{n+1}}\,R_{b\bar{b}}(s).
\end{equation}
In this case the characteristic momentum and energy scales are 
replaced by $p\sim 2 m_b/\sqrt{n}$ and $E\sim m_b/n$, respectively. 
The requirement of perturbativity puts an upper limit on the 
admissible values of $n$. On the other hand, only the resonance 
contribution to the sum rule is experimentally well known, and $n$ needs 
to be taken large enough to reduce the error from the 
$b\bar{b}$ continuum. When $n\gg 1$, the perturbative expansion of 
$M_n$ in the strong coupling $\alpha_s$ breaks down, because there 
exist terms of the form $(\alpha_s\sqrt{n}\,)^k$ in any order of 
perturbation theory. This suggests a summation of the perturbative 
expansion to all orders in which $\alpha_s\sqrt{n}$ is 
treated as order 1 \cite{VZ87}.

In this letter we analyse the sum rule (\ref{eq1}) at
next-to-next-to-leading order (NNLO) in this resummed perturbative
expansion. [A preliminary analysis was presented in
Ref.~\cite{BSS98II}.] The resummed perturbative $b\bar{b}$ cross
section is computed at NNLO using recent 2-loop results on the
Coulomb potential \cite{Sch98} and the $\gamma^* b\bar{b}$ vertex
\cite{BS98,CM98,Hoa97} and non-relativistic effective field theory in
dimensional regularization as described in
\cite{BSS98II,Ben98a,BSS99}.  Rather than determining the $b$ quark
pole mass, as has usually been done, we apply the potential
subtraction (PS) scheme and determine the PS mass \cite{Ben98} from
the sum rule. We expect perturbative corrections in this and related
schemes to be smaller than in the on-shell scheme
\cite{Ben98,HSSW98}. We then convert the extracted PS mass to the
$\overline{\rm MS}$ mass, thus by-passing the infrared sensitivity
problem of the on-shell scheme \cite{BB94}, and yet
implementing the resummation necessary in the non-relativistic
kinematics enforced by taking large moments. Other NNLO analyses of
the sum rule have already appeared \cite{PP98,Hoa98,MY98,Hoa99}.
Nevertheless, we think that an independent analysis, together with a
critical discussion of the quark mass error, is still useful.  We also
perform a complementary analysis and determine the $b$ quark mass
directly from the mass of the $\Upsilon(\mbox{1S})$ state. This has
been done previously in a NNLO analysis presented in \cite{PY98},
which, however, concentrated on the $b$ quark pole mass, as did
\cite{PP98,Hoa98}.

\vspace*{0.2cm}
{\em Experimental moments.} We first evaluate the integrals (\ref{eq1}) 
by expressing the cross section in terms of the six $\Upsilon$ 
resonances and the open $b\bar{b}$ continuum. The masses and leptonic 
widths of the resonances are taken from \cite{pdg98}. Very little 
information exists on the $b\bar{b}$ continuum above 
$s=(10.56\,\mbox{GeV})^2$ \cite{cleo}. We parametrize the continuum 
by setting $R_{b\bar{b}}^{cont.} = 0.4\pm 0.2$. With this crude 
parametrization the experimental error on the determination of $m_b$ 
is $110\,(50,30,15)\,$MeV for $n=4\,(6,8,12)$, and small compared to the 
theoretical error for interesting moments with $n\sim 8$-$12$. 
Some experimental moments are shown in Table~\ref{tab1}. For 
$n= 8$-$12$ about 70\%-85\% of the experimental moment comes from 
the $\Upsilon(\mbox{1S})$ resonance.
\begin{table}[t]
\addtolength{\arraycolsep}{0.1cm}
\renewcommand{\arraystretch}{1.4}
$$
{\small
\begin{array}{c|c|c|c|c|c}
\hline\hline
n & 4 & 8 & 10 & 12 & 16  \\ 
\hline 
M_n  & 0.231\pm 0.044 & 0.235\pm 0.023 & 0.264\pm 0.021 & 
0.308\pm 0.021 & 0.443\pm 0.025   \\
\hline\hline
\end{array}
}
\vspace*{-0.2cm}
$$
\caption[dummy]{\label{tab1}\small The experimental moments.}
\vspace*{-0.2cm}
\end{table}

\vspace*{0.2cm}
{\em Theoretical moments.} The theoretical moments are computed by 
first matching QCD to non-relativistic QCD. In a second step this theory 
is matched to a non-local Schr\"odinger field theory, in which $b\bar{b}$ 
pairs propagate through the Coulomb Green function. We then solve the 
Schr\"odinger equation to NNLO. We refer to \cite{BSS98II,Ben98a,BSS99} 
for some details of the method; further useful information can be found 
in \cite{PP98,Hoa98,MY98}. The result for the $b\bar{b}$ cross 
section to NNLO, still in the on-shell scheme, is expressed as 
\begin{equation}
\label{rr}
R_{b\bar{b}}(s) = \alpha_s \left\{f_0(\lambda,l) + 
\alpha_s \,f_1(\lambda,l)+ 
\alpha_s^2 \,f_2(\lambda,l)\right\},
\end{equation}
where $l=\ln(-4 m_b E/\mu^2)$, $\lambda=C_F\alpha_s/(2\,(-E/m_b)^{1/2})$, 
$C_F=4/3$, $E=\sqrt{s}- 2 m_b$ and $m_b$ is the $b$ pole mass. The functions 
$f_i(\lambda,l)$ contain bound-state poles that correspond to the 
$\Upsilon(\mbox{nS})$ resonances. We obtained these functions analytically. 
After integrating numerically over $s$ according to (\ref{eq1}), these 
functions sum all terms of the form 
$\alpha_s^{0,1,2}\,(\alpha_s\sqrt{n})^k$ to all orders.
Writing $R_{b\bar{b}}(s)$ in the form 
of (\ref{rr}) implies that we expand the bound-state pole $\delta$-functions 
around the leading-order pole position. Expanding the bound-state 
pole $\delta$-functions rather than leaving them unexpanded is motivated by 
the fact that the sum rule relies on global duality. Using dispersion 
relations, the moments can be expressed in terms of derivatives of the 
vacuum polarization as indicated in (\ref{eq1}), which makes no 
reference to individual resonances. Computing these derivatives in 
resummed perturbation theory to NNLO implies that we expand the resonance  
$\delta$-functions in the expression for $R_{b\bar{b}}(s)$.\footnote{
For very large $n$ one should keep the $\delta$-functions unexpanded, 
because the effective smearing interval in $s$ becomes smaller than 
the perturbative correction to the bound-state pole position. 
But for such large $n$ one has to rely on local duality and the sum 
rules suffers from non-perturbative uncertainties as we 
discuss further below.} 

Before integrating over $s$, we convert the expression for 
$R_{b\bar{b}}(s)$ from the on-shell to the potential subtraction scheme. 
The pole mass is eliminated using the relation \cite{Ben98} 
\begin{eqnarray}
\label{mass}
m_b &=& m_{b,\rm PS}(\mu_f) -\frac{1}{2}\int\limits_{|\vec{q}\,|<\mu_f} 
\!\!\!\frac{d^3\vec{q}}{(2\pi)^3}\,\tilde{V}(q) 
\nonumber\\[-0.2cm]
&=& m_{b,\rm PS}(\mu_f) + \frac{C_F\alpha_s}{\pi}\,\mu_f 
\left[1+\frac{\alpha_s}{4\pi}\,\delta m_1 + 
\left(\frac{\alpha_s}{4\pi}\right)^{\!2}\,\delta m_2+\ldots\right],
\end{eqnarray}
where $\tilde{V}(q)$ is the Coulomb potential in momentum space.
Explicit expressions for $\delta m_{1,2}$ can be found in \cite{Ben98}. 
Note that $m_b-m_{b,\rm PS}(\mu_f)$ is proportional to a subtraction 
scale $\mu_f$, which should not exceed the characteristic scale 
$2 m_b/\sqrt{n}$ of the moments $M_n$. We insert 
(\ref{mass}) into (\ref{rr}) and expand the small correction terms 
involving $\delta m_{1,2}$. However, the term 
$\Delta=C_F\alpha_s\mu_f/\pi$ is 
not expanded when $m_b$ is replaced in $E$, $\lambda$ or $l$, because 
$\Delta$ counts as being of the same order as 
$E=\sqrt{s}-2 m_b$. The result is an expression of the same form 
as (\ref{rr}), but with $m_{b,\rm PS}(\mu_f)$ as input parameter. 
As mentioned in the introduction we expect the expansion (\ref{rr}) 
in this new variable to be more 
convergent, and hence the PS mass can be determined 
with smaller error than the pole mass. The PS mass depends on $\mu_f$ 
and we choose $\mu_f=2\,$GeV as our default. The PS masses for 
different $\mu_f$ are connected by a renormalization group equation 
that follows directly from the definition (\ref{mass}).

The dominant theoretical uncertainty arises from the 
residual dependence of the theoretical moments on the renormalization 
scale $\mu$ of the strong coupling $\alpha_s\equiv\alpha_s(\mu)$ in 
the $\overline{\rm MS}$ scheme. We now discuss the choice and variation 
of this scale and the choice of moments $n$ that go into our analysis. 

As indicated in (\ref{rr}) explicit logarithms of $\mu$ always come 
as $\ln(-4 m_b E/\mu^2)$. When $n\gg 1$, the integral (\ref{eq1}) falls 
exponentially as $\exp(-n E/m_b)$, so that the characteristic energy scale 
is $E\sim m_b/n$. This determines the parametric form of the 
renormalization scale to be $\mu_n\equiv 2 m_b/\sqrt{n}$. This, however, 
is not strictly true, because the moments also contain parts 
in which gluons carry momentum of order $m_b$ and momentum of 
order $m_b/n$. In the renormalization-group-improved treatment 
(see \cite{BSS98II,BSS99}) the hard scale $m_b$ enters as the 
starting point $\mu_h\sim m_b$ of the renormalization group evolution 
of the Wilson coefficient functions. The dependence on $\mu_h$ is 
negligible, of the order of $\pm 10\,$MeV on the output for $m_b$, 
compared to the dependence on the scale $\mu_n$, which determines 
the endpoint of the renormalization-group evolution. It is therefore not 
considered further. Gluons with three-momentum of order $m_b/n$ enter 
only at order $\alpha_s^3$. This leaves us with the scale $\mu_n$ 
above and we adopt $\mu_n\equiv 2 m_b/\sqrt{n}$ as the most `natural' 
scale. 

One may object that the form of the logarithm gives the natural 
scale only parametrically, but that the scale 
is arbitrary up to a multiplicative factor, 
since the physical scale in the 
$\overline{\rm MS}$ scheme corresponds to a different scale in another 
scheme, for instance MS. We can address this question by searching 
for constants that appear systematically in conjunction with the logarithm 
$\ln(-4 m_b E/\mu^2)$. While this is complicated for the full cross 
section, it is easily done for the bound state energies that correspond 
to the $\Upsilon(\mbox{nS})$ resonances. We find that for the 
$k$th energy level the analogous 
logarithm always appears in the combination $\ln(m_b C_F\alpha_s/(k \mu))
-S_1(k)$, where $S_1(k)\equiv\sum_{m=1}^k 1/m$. Since $S_1(k)>0$, this 
suggests -- if anything --, that the physical scale is even smaller than 
what we inferred from the logarithm alone.

The useful moments are restricted from below by the uncertainty in the 
experimental value of the moment. If we aim at an error of about 
$50\,$MeV in the determination of the $b$ quark mass, we need $n\geq 6$. 
There is also a technical restriction, which could be overcome. Our 
expression for $M_n$ sums all terms of the form $\alpha_s^{0,1,2} 
(\alpha_s\sqrt{n})^k$, but it does not make use of the exact 
fixed-order coefficients at order $\alpha_s^{1,2}$, which are known 
\cite{CKS97}, because terms of relative order $n^{-3/2}$ or smaller 
are dropped. This could be compensated for by matching the resummed 
result and the fixed-order result. However, we find that for $n\geq 6$ 
this matching correction is small, as can be seen from Table~\ref{tab2}.  

It is advantageous to take large moments, because large moments are more 
sensitive to $m_b$, while the experimental error does not increase, see
Table~\ref{tab1}. An upper limit arises, because the characteristic scales 
must remain perturbative. As concerns $\mu_n$, the requirement 
$\mu_n>\Lambda_{\rm QCD}$ does not seem to pose a serious 
restriction. In practice, we find that the theoretical prediction 
becomes unstable already when $\mu$ is smaller than 1.5-2.0$\,$GeV;  
requiring $\mu$ to be larger than this is restrictive, if we also allow 
for a variation of $\mu$ about $\mu_n$. A more serious constraint 
arises from the scale $m_b/n$, which enters the NNLO calculation 
implicitly. At NNNLO there is a contribution to the moment that scales as 
$\alpha_s(\mu_n)^2\alpha_s(m_b/n)$.  
When $m_b/n \sim \Lambda_{\rm QCD}$, 
we should count $\alpha_s(m_b/n)$ as order 1. In this case, we have 
an uncontrolled non-perturbative contribution to the moments that is 
formally of NNLO.\footnote{It is worth noting that 
the scales $\mu_n$ and $m_b/n$ do not really approach 0 as $n\to \infty$, 
but freeze at values of order $m_b \alpha_s$ and $m_b\alpha_s^2$, 
respectively. While this is of interest for a very heavy quark, it 
is of little practical relevance to $b$ quarks.} 
We therefore require $n\leq 10$. In the literature 
larger moments are often used. 
The justification for this is that 
the gluon condensate contribution to the moments, which represents the 
leading non-perturbative power correction, is small even for moments 
much larger than 10. However, the operator product expansion in local 
operators is itself only valid when $m_b/n > \Lambda_{\rm QCD}$ and 
so the estimate is not rigorous. It may, however, indicate that ultrasoft 
contributions from the scale $m_b/n$ are smaller than what we would 
estimate on parametric grounds.

For reference we give some selected moments in Table~\ref{tab2} in the 
on-shell and PS scheme. The table also quantifies the importance of 
resumming $(\alpha_s\sqrt{n})^k$ corrections and the error incurred by 
not including the exact fixed-order coefficients at order $\alpha_s^{1,2}$. 
Resummation is crucial even for $n=6$, because the contribution from the 
bound-state poles, which does not exist in the NNLO fixed-order 
approximation, is large. On the other hand, already for $n=6$ the 
fixed-order moments (FO1) are well approximated by the leading three 
terms in their large-$\sqrt{n}$ expansion (FO2).

\begin{table}[t]
\addtolength{\arraycolsep}{0.1cm}
\renewcommand{\arraystretch}{1.4}
$$
{\small
\begin{array}{c||c|c|c||c|c|c||c|c|c}
\hline\hline
&\multicolumn{3}{c||}{\mbox{LO}}&\multicolumn{3}{c||}{\mbox{NLO}}&
\multicolumn{3}{c}{\mbox{NNLO}} \\
\hline
n & \mbox{Res.} & \mbox{FO1} & \mbox{FO2} & \mbox{Res.} & 
\mbox{FO1} & \mbox{FO2} & \mbox{Res.} & \mbox{FO1} & \mbox{FO2}  \\ 
\hline 
6  & 0.134 \, [0.228] & 0.030 & 0.028 
   & 0.127 \, [0.143] & 0.062 & 0.065
   & 0.201 \, [0.222] & 0.094 & 0.097 \\
\hline 
10 & 0.122 \, [0.268] & 0.016 & 0.016 
   & 0.143 \, [0.166] & 0.041 & 0.042
   & 0.266 \, [0.322] & 0.073 & 0.075 \\
\hline 
14 & 0.139 \, [0.379] & 0.011 & 0.011
   & 0.190 \, [0.218] & 0.032 & 0.033  
   & 0.400 \, [0.529] & 0.065 & 0.067 \\
\hline\hline
\end{array}
}
\vspace*{-0.2cm}
$$
\caption[dummy]{\label{tab2}\small Selected moments (recall the 
normalization of the moments in (\ref{eq1})) in the on-shell and 
PS scheme (square brackets) for $\alpha_s(M_Z)=0.118$ and $m_b=4.95\,$ 
GeV (on-shell scheme) and $m_{b,\rm PS}(2\,\mbox{GeV})=4.55\,$GeV 
(PS scheme). The renormalization scale is taken to be $\mu_n=2 M/\sqrt{n}$, 
where $M=m_b$ or $M=m_{b,\rm PS}(2\,\mbox{GeV})$. 
`Res.' refers to the resummed cross section with LO, NLO, 
NNLO in the sense of (\ref{rr}). `FO1' refers to the fixed-order 
result without resummation, including terms of order $\alpha_s^{0,1,2}$ 
(LO, NLO, NNLO). `FO2' refers to the fixed-order result, dropping terms 
with relative suppression $n^{-3/2}$ or more. 
The effect of resummation is given by the 
difference of `Res.' and `FO2'. The correction due to matching to the 
fixed-order result is roughly given by `FO1' minus `FO2'.}
\vspace*{-0.2cm}
\end{table}

\begin{figure}[t]
   \vspace{-2.3cm}
   \epsfysize=11cm
   \epsfxsize=8cm
   \centerline{\epsffile{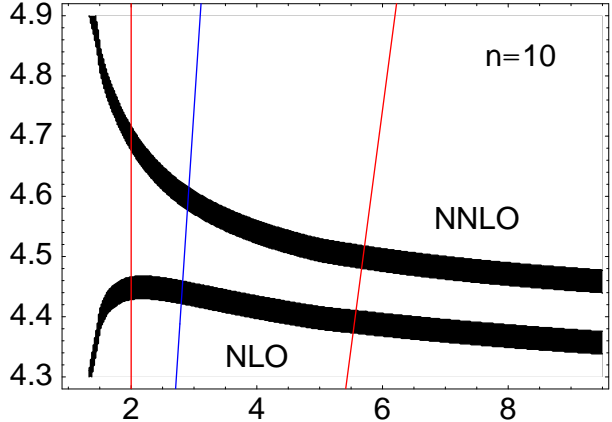}}
   \vspace*{-2.9cm}
\caption[dummy]{\label{mbfig1} \small 
The value of $m_{b,\rm PS}(2\,\mbox{GeV})$ 
obtained from the 10th moment as a function of the renormalization scale 
in NLO and NNLO and for $\alpha_s(m_Z)=0.118$. 
The dark region specifies the variation due to the 
experimental error on the moment. The middle line marks the scale 
$\mu_n$, the two outer lines determine the scale variation from which the 
theoretical error is computed.}
\end{figure}

\vspace*{0.2cm}
{\em Numerical analysis.} For a given $n$ the theoretical moments are 
functions of $m_{b,\rm PS}(\mu_f)$, which we would like to determine; 
the strong coupling $\alpha_s$, for which we use 
$\alpha_s(m_Z)=0.118\pm 0.003$ together with 3-loop evolution; and 
the renormalization scale $\mu$, which is our (rough) handle to estimate 
the uncertainty due to the NNLO approximation of the moment calculation. 
We first compute, for given $n$ and $\mu$ (and $\alpha_s(m_Z)=0.118$), 
the values of $m_{b,\rm PS}(2\,\mbox{GeV})$, 
for which the theoretical moment lies within the 
experimental range. For $n=10$ the result is shown in Fig.~\ref{mbfig1}. 
It is evident that the resulting $m_{b,\rm PS}(2\,\mbox{GeV})$ 
varies significantly as function of 
the scale $\mu$ at which the sum rule is evaluated. Furthermore, there is 
no overlap between the range of masses that is 
obtained from the NNLO and the 
NLO sum rule for any reasonable range of $\mu$. The same conclusion is 
obtained for $n=6$ or $8$. We also determined $m_{b,\rm PS}(2\,\mbox{GeV})$ 
from $n=6,8,10$ 
simultaneously by minimizing a $\chi^2$ with equal weights. The value 
we obtain from this procedure differs by no more than $30\,$MeV 
from that obtained from single moments, when $\mu$ is varied between 
$1.5$ and $9.5\,$GeV, reflecting that the $\mu$-dependence of the 
theoretical moments is completely correlated. The same analysis in the 
on-shell scheme results in an identical qualitative picture; however, the 
scale-dependence is even larger in the on-shell scheme.

As explained above, we take $\mu_n=2 m_{b,\rm PS}(\mu_f)/\sqrt{n}$ as our 
default choice of scale. We would then follow common practice and estimate 
a theoretical error by varying the scale between one half and twice this 
value. But from Fig.~\ref{mbfig1} we observe that the theoretical 
prediction becomes unstable (compare the behaviour of the NLO and NNLO 
results) for scales below $2\,$GeV and one may argue that varying the 
scale into this region does not provide a reliable error estimate. We 
therefore compute the theoretical error from a variation between $2\,$GeV 
and $2\mu_n$. It is clear that the error so estimated is rather sensitive 
to the lower scale cut-off. Taking $n=10$ and adding 
the error from $\alpha_s$ and the experimental moments, 
we obtain
\begin{equation}
\label{psmass}
m_{b,\rm PS}(2\,\mbox{GeV}) = 
(4.60\pm 0.10\,(\mbox{scale})\pm 0.03\,(\alpha_s) 
\pm 0.02\,(\mbox{exp.}))\,\mbox{GeV}.
\end{equation}
If the scale is varied down to $2.5 \,(1.5)\,$GeV, the scale error decreases 
(increases) to $\pm\,65 \,(160)\,$MeV. In comparison, a NLO analysis of 
the sum rule would return the central value 
$m_{b,\rm PS}(2\,\mbox{GeV}) = 4.44\,$GeV with 
a smaller scale uncertainty (see Fig.~\ref{mbfig1}). The large difference 
with the NNLO result casts doubt on the convergence of successive perturbative 
approximations. The origin of this difference and the origin of the large 
scale dependence will become clear below.

The PS mass is a useful parameter (replacing the pole mass) for 
short-distance observables involving $b$ quarks close to their mass shell. 
For high energy processes, we would like to convert the PS mass to 
the $\overline{\rm MS}$ definition. Call $\overline{m}_b$ the 
$\overline{\rm MS}$ $b$ quark mass at the renormalization scale 
$\overline{m}_b$ and $k_r$ the coefficient at order 
$(\alpha_s(\overline{m}_b)/(4\pi))^r$ that relates the pole 
mass to $\overline{m}_b$. From (\ref{mass}) we obtain the relation 
\begin{equation}
\label{msrel}
m_{b,\rm PS}(\mu_f) = \overline{m}_b\left(1+\sum_{r=1}^\infty 
\bigg[k_r-4 C_F\delta m_{r-1}(\mu_f)\,\frac{\mu_f}{\overline{m}_b}\,\bigg] 
\,\frac{\alpha_s(\overline{m}_b)^r}{(4\pi)^r}\right),
\end{equation}
where we defined $\delta m_0\equiv 1$. An N$^k$LO analysis of the sum 
rule determines the PS mass with a {\em parametric} accuracy of order 
$m_{b,\rm PS}\,\alpha_s^{k+2}$. This is most easily seen by noting that 
an N$^k$LO calculation of the resummed $b\bar{b}$ cross section determines 
the (perturbative) $\Upsilon(\mbox{nS})$ masses to order 
$m_{b,\rm PS}\alpha_s^{k+2}$. To determine the $\overline{\rm MS}$ mass 
with the same parametric accuracy implies that one should use (\ref{msrel}) 
at order $\alpha_s^{k+2}$. At present (\ref{msrel}) is known only 
to third order, combining the result of \cite{CS99} for $k_3$ and the 
one for $\delta m_2(\mu_f)$ from \cite{Ben98}. 

To obtain an order-of-magnitude estimate of the missing fourth-order 
term, we estimate the coefficient in the `large-$\beta_0$' approximation 
\cite{BBB,Neu95}. This gives $k_4=339071$ and  
$\delta m_3(2\,\mbox{GeV})=125026$ for $n_f=4$ and 
$\overline{m}_b=4.26\,$GeV.\footnote{For comparison, note that the 
`large-$\beta_0$' approximation for $\delta m_2(2\,\mbox{GeV})$  
results in 2140.36 rather than the exact value of 1870.54. For $k_3$ 
one obtains 6526.91 rather than the `exact' value 6144(128). (The brackets 
specify the error on the `exact' result, see \cite{CS99}.) $n_f$ refers to 
the number of light-quark flavours.} 
Although the individual coefficients are large, there are large 
cancellations in the combination that enters (\ref{msrel}), which 
reflect the infrared cancellation that motivated the introduction of 
the potential subtraction \cite{Ben98}. With these numbers, given 
$m_{b,\rm PS}(2\,\mbox{GeV})$, we estimate that the $\alpha_s^4$ term 
reduces $\overline{m}_b$ by $10\,$MeV. (For comparison, the $\alpha_s^3$ 
term provides a $35\,$MeV reduction.) 
We therefore assume an additional 
$(10\pm 20)\,$MeV correction in the relation between 
$\overline{m}_b$ and $m_{b,\rm PS}(2\,\mbox{GeV})$ beyond the 
third-order formula. This results in the $\overline{\rm MS}$ mass\footnote{
We compute $\overline{m}_b$ by solving (\ref{msrel}) exactly for a given 
PS mass, rather then inverting (\ref{msrel}) perturbatively to order 
$\alpha_s^4$. The second procedure would result in a central value 
that is negligibly different by $4\,$MeV from the one given.} 
\begin{equation}
\label{msbar}
\overline{m}_b = (4.26\pm 0.09\,(\mbox{scale})\pm 0.01\,(\alpha_s) 
\pm 0.02\,(\mbox{conv.})\pm 0.02\,(\mbox{exp.}))\,\mbox{GeV}.
\end{equation}
The dependence on $\alpha_s$ nearly cancels out and `conv.' refers to 
the conversion from the PS to the $\overline{\rm MS}$ scheme just 
discussed. We have repeated the analysis with $\mu_f=1\,$GeV 
and $3\,$GeV for 
the subtraction scale of the PS mass. Converting to $\overline{m}_b$, 
we find agreement with (\ref{msbar}) within $20\,$MeV.

\begin{figure}[t]
   \vspace{-2.6cm}
   \epsfysize=11cm
   \epsfxsize=8cm
   \centerline{\epsffile{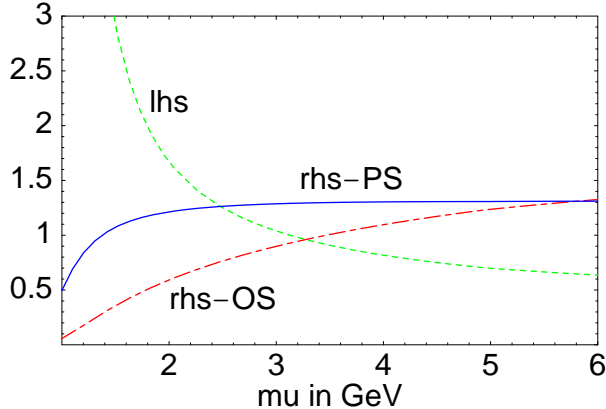}}
   \vspace*{-2.9cm}
\caption[dummy]{\label{mbfig2} \small The left-hand side (lhs, 
short-dashed line) and right-hand side (rhs, dash-dotted and solid lines) of 
(\ref{truncated}) as a function of the scale $\mu$ for $n=10$, 
$m_b=5\,$GeV (on-shell scheme, dash-dotted) and 
$m_{b,\rm PS}(2\,\mbox{GeV})=4.6\,$GeV (PS scheme, solid). The figure 
clearly shows the reduction of the scale dependence in the PS 
scheme for the predicted $\Upsilon({\rm 1S})$ mass. The scale dependence of 
the width (short-dashed line) is identical in the on-shell and PS schemes.}
\end{figure}

\vspace*{0.2cm}
{\em Origin of the large scale dependence.} The scale uncertainty in 
(\ref{psmass}) is only about 30\% smaller than the uncertainty we would 
have found in the on-shell scheme. To understand the origin of this 
marginal improvement, we consider a truncated sum rule, in which both  
experimental and theoretical moments  are given only in terms of the 
first $\Upsilon$ resonance. This is actually not a bad approximation 
to the full sum rule and allows us to discuss the origin of scale dependence 
in a transparent form. In this approximation, performing in addition a 
non-relativistic approximation to the $s$-integration measure in 
(\ref{eq1}), we can write the sum rule as
\begin{equation}
\label{truncated}
\frac{\Gamma^{th}_{l^+ l^-}}{\Gamma^{exp}_{l^+ l^-}} = 
\exp\bigg[-(2 n+1)\,\frac{M_{\Upsilon({\rm 1S})}^{exp}-
M_{\Upsilon({\rm 1S})}^{th}}{M_{\Upsilon({\rm 1S})}^{exp}}\bigg],
\end{equation}
where $\Gamma^{th}_{l^+ l^-}$ and $M_{\Upsilon({\rm 1S})}^{th}$ are the 
leptonic width and mass of the $\Upsilon(\mbox{1S})$ state computed to 
NNLO. In the on-shell scheme, the series expansions for the two quantities 
read\footnote{The mass and leptonic width have been obtained 
to NNLO in \cite{PY98,MY98} and \cite{MY98}, respectively. Our analytic 
expressions for an arbitrary $\Upsilon(\mbox{nS})$ state coincide with 
those previous results, provided we neglect the renormalization-group 
improvement for the leptonic width. At present only the logarithms from the 
renormalization of the external current are taken into account.}
\begin{eqnarray}
M_{\Upsilon({\rm 1S})}^{th} &=& 2 m_b - \frac{4}{9}\,m_b\alpha_s^2
\,\bigg(1+ \big[3.590-1.326 \,l\big]\,\alpha_s + \big[19.52-6.033 \,l+
1.319 \,l^2\big]\,\alpha_s^2\bigg),
\nonumber\\
&=& 2 m_b - \frac{4}{9}\,m_b\alpha_s^2
\,\bigg(1+1.08+1.76+\ldots\bigg)
\label{mups}\\ 
\Gamma^{th}_{l^+ l^-} &=& \frac{32}{27}\,e_b^2\alpha_{em}^2 m_b 
\alpha_s^3\,
\bigg(1+\big[3.605-1.989\,l-5.607\,\kappa\big]\,\alpha_s 
\nonumber\\
&&\hspace*{-1cm}
+ \,\big[24.14-7.898\,l+2.639\,l^2-0.3793\,\ln\kappa-20.21\kappa+
11.16\,l\kappa+7.260\,\kappa^2\big]\,\alpha_s^2\bigg),
\nonumber\\
&=& \frac{32}{27}\,e_b^2\alpha_{em}^2 m_b 
\alpha_s^3\,
\bigg(1-0.11+1.23+\ldots\bigg)
\label{width}
\end{eqnarray}
where $\alpha_s=\alpha_s(\mu)$, $l=\ln(16 m_b^2\alpha_s^2/(9\mu^2))$ and 
$\kappa=\alpha_s(m)/\alpha_s(\mu)$ and the second line is given for 
$\mu$ such that $l=0$ (for $m_b=5\,$GeV), in which case 
$\alpha_s(\mu)=0.30$. Neither of the series is converging and, because of 
the large power of $\alpha_s$ of the NNLO term, the scale dependence 
is huge at small scales, where $\alpha_s$ varies fast. This is seen from 
Fig.~\ref{mbfig2}, which shows the scale dependence of the left-hand side 
and right-hand side of (\ref{truncated}) separately.

Going from the on-shell to the PS scheme improves the convergence and 
scale dependence of the predicted $\Upsilon({\rm 1S})$ mass as seen from 
the solid line in Fig.~\ref{mbfig2}, but has little effect on the 
series expansion (\ref{width}) for the width. Hence the large scale 
uncertainty in (\ref{psmass}) can be traced to the poor control over the 
perturbative expansion for the leptonic width that controls the over-all 
normalization of the theoretical moments.

\vspace*{0.2cm}
{\em Constraints from $M_{\Upsilon({\rm 1S})}$.} The fact that the mass 
determination from the sum rule is dependent on the theoretical prediction 
of the leptonic width, and limited in accuracy for this reason, suggests 
that we consider determining $m_b$ directly from the $\Upsilon$ spectrum. 
Non-perturbative corrections to the $\Upsilon$ masses grow rapidly for 
higher radial excitations and preclude using any state other than the 
ground state. The problem with this method is that even for the 
$\Upsilon(\mbox{1S})$ state it is difficult to estimate the 
non-perturbative correction 
reliably. The perturbative expression for the $\Upsilon(\mbox{1S})$ 
mass in the on-shell scheme is given by (\ref{mups}) above.

If $m_b (C_F\alpha_s)^2/4\gg \Lambda_{\rm QCD}$, the non-perturbative 
correction to the $\Upsilon(\mbox{1S})$ mass can be 
computed in terms of vacuum 
condensates of local operators. The leading contribution is \cite{LV81}
\begin{equation}
\label{gluoncond}
\delta M_{\Upsilon({\rm 1S})}^{\rm np} = \frac{624}{425}\,\pi m_b\,
\frac{\langle \alpha_s GG\rangle}{(m_b C_F\alpha_s)^4},
\end{equation}
where $\langle \alpha_s GG\rangle\approx 0.05\,\mbox{GeV}^4$ is the 
gluon condensate. The actual magnitude of 
$\delta M_{\Upsilon({\rm 1S})}^{\rm np}$ is rather uncertain. If we 
choose the `natural' scale $\mu=m_b C_F\alpha_s(\mu)\approx 2\,$GeV, 
we obtain $\delta M_{\Upsilon({\rm 1S})}^{\rm np}\approx 90\,$MeV. 
However, as noted earlier, the logarithm that determines this scale appears 
together with constants that tend to make the effective scale lower. 
Furthermore, it may be argued that the coupling $\alpha_s$ should be 
taken as the coefficient of $C_F/r$ in the Coulomb potential rather than 
in the $\overline{\rm MS}$ scheme. This coupling is larger than the 
$\overline{\rm MS}$ coupling. Both effects can decrease 
$\delta M_{\Upsilon({\rm 1S})}^{\rm np}$ substantially.

Since the inequality $m_b (C_F\alpha_s)^2/4\gg \Lambda_{\rm QCD}$ that 
justifies the operator product expansion (OPE) does not hold, we should 
consider the subsequent term in the OPE to judge whether the expansion 
converges. Using the result of \cite{Pin97}, we find that the 
contribution from dimension-6 operators could be anything between 
a fraction of and twice $-\delta M_{\Upsilon({\rm 1S})}^{\rm np}$, where the 
large uncertainty stems from the poorly known dimension-6 condensates 
and the ambiguity in the value of $\alpha_s$.\footnote{Ref.~\cite{Pin97} 
concludes that the OPE appears to be convergent, because the minimal 
value of the strong coupling in the denominator of (\ref{gluoncond}) 
assumed there is larger than the minimal value 
allowed in our estimate.} This puts the convergence of the OPE in question. 
We therefore consider (\ref{gluoncond}) as an order-of-magnitude estimate 
of the non-perturbative correction and treat it as a theoretical error 
rather than adding it to (\ref{mups}). In our opinion, assigning an 
error of $\pm 70\,$MeV to $m_b$ from this source is conservative.

We proceed to determine $m_{b,\rm PS}(\mu_f)$ from $M_{\Upsilon({\rm 1S})}$ 
using (\ref{mups}) converted to the PS scheme. This renders the series 
(\ref{mups}) convergent and leads to a very small scale uncertainty of 
the extracted value of $m_{b,\rm PS}(2\,\mbox{GeV})$,
as anticipated from the solid curve in Fig.~\ref{mbfig2}. 
Varying $\mu$ from $1.25\,$ to $4\,$GeV, we obtain
\begin{equation}
\label{psmass2}
m_{b,\rm PS}(2\,\mbox{GeV}) = 
(4.58\pm 0.04\,(\mbox{scale})\pm 0.01\,(\alpha_s) 
\pm 0.07\,(\mbox{non-pert.}))\,\mbox{GeV},
\end{equation}
which is consistent with (\ref{psmass}). 
In this case a NLO analysis would return 
the central value $m_{b,\rm PS}(2\,\mbox{GeV})=4.53\,$GeV, which suggests 
that the corresponding small value in the sum rule analysis is an anomaly 
related to the behaviour of the series for the leptonic width. 
From (\ref{psmass2}) we obtain the $\overline{\rm MS}$ mass 
\begin{equation}
\label{msbar2}
\overline{m}_b = (4.24\pm 0.04\,(\mbox{scale},\mu_f)\pm 0.02\,(\alpha_s) 
\pm 0.02\,(\mbox{conv.})\pm 0.07\,(\mbox{non-pert.}))\,\mbox{GeV}.
\end{equation}
The central value varies by only about $10\,$MeV, when $\mu_f$ is 
varied between $1$ and $3\,$GeV. Contrary to the 
sum rule determination, the error is dominated by the non-perturbative 
contribution to the $\Upsilon(\mbox{1S})$ mass. This leaves room for 
improving upon the error, if some quantitative insight into the 
non-perturbative contribution could be obtained.

\vspace*{0.2cm} {\em Comparison with previous results.} We compare the
bottom quark mass obtained in this work with the results of earlier
NNLO analyses of the sum rule and the $\Upsilon(\mbox{1S})$ mass. Our
comments will be restricted to those analyses that quote a result for
the $\overline{\rm MS}$ quark mass \cite{PP98,Hoa98,MY98,Hoa99,PY98,JP98}.
With the exception of \cite{PY98}, which obtains $\overline{m}_b$
from $M_{\Upsilon({\rm 1S})}$ and, therefore, should be compared with
(\ref{msbar2}), all other NNLO analyses use the sum rule (\ref{eq1})
and should be compared with (\ref{msbar}).

The value given in (\ref{msbar2}) is significantly smaller than
$\overline{m}_b = (4.44\pm 0.04)\,$GeV,  obtained by \cite{PY98}. This
difference is explained by the fact that \cite{PY98} first uses the on-shell
scheme to extract the pole mass and then uses the 2-loop
truncation of (\ref{msrel}) [with $\mu_f=0$] to obtain
$\overline{m}_b$. However, contrary to the PS scheme with
$\mu_f=2\,$GeV, the 3-loop and 4-loop terms are large in the on-shell
scheme; at least the 3-loop term\footnote{Since the large infrared 
contribution in the 4-loop term cancels against an NNNLO contribution 
to the $\Upsilon(\mbox{1S})$ mass, it can be argued that only the 3-loop term 
is to be used. This is different from the PS scheme, where no 
systematically large coefficients appear. Compare (\ref{msrel}) [with 
$\mu_f\sim m_b\alpha_s$] and the discussion in \cite{Ben98, HLM99} 
regarding combining different powers of $\alpha_s$ to make infrared 
cancellations manifest.} has to be included when the pole 
mass is determined from
the NNLO formula for the $\Upsilon(\mbox{1S})$ mass. Estimating the
terms missing in \cite{PY98} in the large-$\beta_0$ limit, and
subtracting them from $\overline{m}_b = (4.44\pm 0.04)\,$GeV, we find
that the result of \cite{PY98} becomes (roughly) consistent with ours. 
The error estimate of \cite{PY98} is, however, less conservative than ours.

A related difficulty concerns the comparison with the value 
$\overline{m}_b = (4.21\pm 0.11)\,$GeV quoted in \cite{PP98}. While 
apparently consistent with the one obtained in this work, it is obtained 
via a 2-loop relation from the $b$ quark pole mass, which in turn 
is determined from the sum rule. If we add the 3-loop and/or 4-loop term, 
the result of \cite{PP98} would be about $200\,$MeV lower than ours. 
This difference is a reflection of the fact that the pole mass quoted 
in \cite{PP98} is roughly $200\,$MeV lower than the one we would have 
obtained had we chosen to determine it. This difference in turn can be 
traced to the use of a high renormalization scale for the evaluation 
of the sum rule, cf. Fig.~\ref{mbfig1}. In our opinion, the choice of 
such a high scale is not well motivated. We also think that it is mandatory 
to use intermediate mass definitions such as the PS mass to determine 
$\overline{m}_b$ reliably. Otherwise large perturbative coefficients make 
it impossible to disentangle true theoretical errors from correlated and 
spurious ones caused by those large perturbative coefficients.

The most recent analysis by the authors of \cite{JP98} determines the
$\overline{\rm MS}$ mass directly from the sum rule and gives
$\overline{m}_b=4.19\pm 0.06\,$GeV from moments with $n=7$-$15$. No
resummation is performed, because it is assumed, incorrectly, that this
is unnecessary in the $\overline{\rm MS}$ scheme. However, for high
moments the Coulomb interaction must be treated non-perturbatively,
and a resummation has to be done, irrespective of the mass renormalization
convention. The $\overline{\rm MS}$ scheme actually makes the
expansion worse, because the expansion contains terms of order
$(\alpha_s n)^k$ in addition to $(\alpha_s \sqrt{n})^k$. To avoid such
terms, one has to use an intermediate convention, such as the potential
subtraction scheme, and then relate this convention to the
$\overline{\rm MS}$ scheme in a second step. Because of this
theoretical shortcoming, the result of \cite{JP98} cannot be compared
with (\ref{msbar}).

The other papers quoted above perform a NNLO resummation as in this
work. Differences arise either in the representation of the
NNLO-resummed moments or in the analysis and error evaluation
strategy. Both \cite{MY98} (MY) and \cite{Hoa99} (Hoang) also use
intermediate mass subtractions, different from the PS scheme, 
but conceptually
similar to it, before converting these intermediate masses
to the $\overline{\rm MS}$ scheme.\footnote{As the analysis in
\cite{Hoa99} supersedes \cite{Hoa98}, we do not discuss \cite{Hoa98}
in detail.}

The differences in the theoretical representation of the moments are 
the following: 
\begin{itemize}
\item[(a)] MY and Hoang use a factorization scheme different from
dimensional regularization. Since the final result is physical, this
is a technical difference that should bear no consequences on the
final result.
\item[(b)] Hoang applies the non-relativistic approximation  
also to the $s$-integration in (\ref{eq1}), while MY and the 
present work obtain the resummed cross section analytically and 
then integrate it numerically according to (\ref{eq1}) or after an 
equivalent contour deformation into the complex $s$-plane. The difference 
is negligible. 
\item[(c)] MY and Hoang have taken the short-distance coefficient 
as an over-all factor, while we have multiplied it out 
to NNLO. Keeping it as an over-all factor is problematic, because this  
results in a spurious factorization scheme and scale dependence, which 
is not small as can be seen from Table~3 of \cite{Hoa99}. Since both  
short-distance and long-distance contributions are computed perturbatively, 
the factorization scale is a purely technical construct and no dependence 
on it should be left in the result. One motivation for writing the 
short-distance coefficient as an over-all factor is that the scales 
in the coupling constant are different in the long- and short-distance 
parts. However, this effect, related to logarithms of $n$, can be treated 
consistently only in the context of a full renormalization group 
treatment. This has been done in the present work (see \cite{BSS98II,BSS99}), 
but not in \cite{MY98,Hoa99}. 
As a consequence there is no analogue of $\mu_{\rm fac}$ in the 
present approach, while the role of $\mu_{\rm hard}$ in \cite{MY98,Hoa99}  
is taken by the starting scale for the renormalization group evolution. 
As mentioned earlier, this dependence is negligible in our representation 
of the moments. 
\item[(d)] Hoang and this work expand the bound-state
$\delta$-functions for reasons discussed earlier; MY keep them
unexpanded.  This increases the theoretical moments significantly at
NNLO. Not expanding the bound state $\delta$-functions would increase
the value quoted in (\ref{psmass}) and (\ref{msbar}) by almost
$100\,$MeV.
\item[(e)] MY and Hoang use a 2-loop formula to obtain 
$\overline{m}_b$ from their intermediate mass. 
As explained above, a NNLO analysis of the sum rule determines
the PS (or related) masses with $m_b \alpha_s^4$ accuracy and to fully
exploit this accuracy, the 4-loop relation between the PS and
$\overline{\rm MS}$ mass should be used. We estimated that the 3- and
4-loop terms decrease $\overline{m}_b$ by $45\,$MeV and this
additional shift has been incorporated in (\ref{msbar}). Employing 
a less accurate relation entails a corresponding loss in parametric 
accuracy of $\overline{m}_b$, although this is a numerically small effect, if 
our estimate is correct.
\end{itemize}

As in this work, MY obtain their result from an analysis of single
moments (checking consistency between a set of moments), although
larger moments $n=14$-$18$ are used, which could be considered
problematic. They use the so-called kinetic mass as
intermediate mass definition. The analogue of $\delta m_2(\mu_f)$ in
(\ref{mass}), needed for a NNLO sum rule analysis, is not yet known in
this scheme; MY estimate it in the large-$\beta_0$ limit, an
additional assumption we had to use only for the 4-loop
term when relating (\ref{psmass}) to (\ref{msbar}), but not to
extract the PS mass in the first place. MY vary the renormalization
scale from $2\,$GeV to $4.5\,$GeV, while we would argue that, for the
high moments used in \cite{MY98}, the scale should be chosen lower. If
we repeat our analysis with the same assumptions as those of MY, we
reproduce their error estimate for the kinetic mass, which is smaller
than the more conservative procedure that leads to (\ref{psmass}).

Hoang uses an analysis and error estimate that is different from the 
single-moment analysis performed by MY and in this work. Simplifying 
somewhat, Hoang fits the quark mass from the linear 
combination $0.12 M_4-0.56 M_6+0.76 M_8-0.31 M_{10}$ of moments, where 
the coefficients are determined by the covariance matrix of the 
{\em experimental} input data such as the measured leptonic widths of the 
six $\Upsilon$ resonances. This linear combination (which entails a 
cancellation of one part in 4000) turns out to be very insensitive to 
the renormalization scale $\mu$, yet retaining a large sensitivity 
to $m_b$. Hoang then scans the theoretical parameter space and 
finds an error of only $\pm 30\,$MeV for the so-called 1S-mass, compared 
to the error of (\ref{psmass}). We have repeated our analysis for this 
linear combination and obtain 
$m_{b,\rm PS}(2\,\mbox{GeV})=4.58\pm 0.02\,$GeV in this way, where 
the quoted error is due to variation of the renormalization scale only. 
The result is consistent with (\ref{psmass}), but the error is much smaller. 
The central value is $50\,$MeV higher than the value reported 
in Sect.~6 of \cite{Hoa99}. Such differences can be explained by different 
implementations of the NNLO result, as discussed above.

Several circumstances make us suspicious that the theoretical error is 
underestimated by Hoang's procedure. For example, if we 
increase the error of the leptonic width of the $\Upsilon(\mbox{2S})$ 
by a factor of 10, or if we increase the error on the measured mass 
of the $\Upsilon(\mbox{1S})$ to $5\,$MeV, which is still small compared 
to the expected theoretical error, the procedure chooses a 
linear combination that exhibits less stability in $\mu$, and has no or 
two solutions for $m_{b,\rm PS}(2\,\mbox{GeV})$ for some ranges 
of $\mu$, even though our experimental knowledge of the width or mass should 
have no bearing on the theoretical error estimate. 
This remark may not be considered as a serious objection, because we 
could abandon the way the linear combination is chosen in \cite{Hoa99} 
and optimize it deliberately. However, even for the 
original linear combination, there is a second solution 
$m_{b,\rm PS}(2\,\mbox{GeV})=4.86\pm 0.04\,$GeV in addition to 
$m_{b,\rm PS}(2\,\mbox{GeV})=4.58\pm 0.02\,$GeV, because the 
linear combination is no longer a monotonic function of $m_b$. The 
criterium of renormalization scale stability does not exclude obtaining 
solutions that differ by more than the error estimated from the 
$\mu$-dependence. The problem is compounded by the observation 
that the stability under variations of the renormalization 
scale, and hence the small error obtained by Hoang, crucially depends on 
the assumption that the four moments are combined at the same value of the 
renormalization scale $\mu$. This is a serious assumption, in particular 
as the natural scale of the moments is $2 m_b/\sqrt{n}$. If we combine 
the moments at their natural rather than at equal scales, the stability is 
lost. To be fair, we should mention that Hoang's analysis is more involved 
than analysing a single linear combination, although the covariance matrix is 
such that it does indeed give most weight to a single one. Nevertheless, 
we think that the simplified discussion above emphasizes the problem 
with estimating a theoretical error in the way done in \cite{Hoa99}.

The final results by MY 
($\overline{m}_b=4.2\pm 0.1\,$GeV) and by Hoang  
($\overline{m}_b=4.20\pm 0.06\,$GeV) agree with (\ref{msbar}) within 
the quoted errors. However, if the $45\,$MeV shift were applied to those 
results, there is actually a discrepancy of about $100\,$MeV in the 
central value. This could be a consequence of the different representations 
of the moments as discussed above. However, we find it difficult 
to reconcile a $\overline{m}_b$ significantly smaller than $4.25\,$GeV 
with the analysis of $M_{\Upsilon({\rm 1S})}$ [see (\ref{msbar2})], 
unless there is indeed a large positive non-perturbative contribution 
to $M_{\Upsilon({\rm 1S})}$.

\vspace*{0.2cm}
{\em Summary.} We determined the bottom quark mass in the 
$\overline{\rm MS}$ scheme and the potential subtraction (PS) scheme 
\cite{Ben98} at next-to-next-to-leading order from sum rules for 
the $b\bar{b}$ cross section and the mass of the $\Upsilon(\mbox{1S})$ 
state. The results are in excellent agreement with each other as 
summarized by (\ref{psmass}), (\ref{msbar}), (\ref{psmass2}) and 
(\ref{msbar2}). There is no systematic procedure to combine the 
two results. On the one hand, the two determinations are not independent, 
because some theoretical input is common to both. On the other hand, 
the dominant source of theoretical error is different. We therefore 
combine the two determinations to yield the PS mass 
\begin{eqnarray}
\label{psmass3}
&&m_{b,\rm PS}(2\,\mbox{GeV}) = 4.59 \pm 0.08\,\mbox{GeV}
\end{eqnarray}
and the $\overline{\rm MS}$ mass (at the scale of the 
$\overline{\rm MS}$ mass)
\begin{eqnarray}
\label{msbar3}
&&\hspace*{-0.8cm}\overline{m}_b(\overline{m}_b) = 
4.25 \pm 0.08\,\mbox{GeV}.
\end{eqnarray}
The calculations that go into these results imply partial resummations 
of the QCD perturbative expansion to all orders. An important point is 
that a NNLO resummation allows us to determine the quark masses with a 
parametric accuracy of order $m_b \alpha_s^4$, i.e. the residual error 
scales formally as $\delta m_b/m_b\sim \alpha_s^5$. In the case of the 
$\overline{\rm MS}$ mass this requires that one controls the four-loop 
relation to the PS mass. We estimated the 4-loop term, which is 
not yet known exactly and found that it should be very small. 

Unfortunately, the sum rule analysis yields a much less precise 
determination of the bottom quark mass than what might have been expected 
with NNLO accuracy. We identify as the reason for this the bad behaviour 
of the perturbative expansion for the leptonic width of the 
$\Upsilon$ resonances. The same is true for the $\Upsilon(\mbox{1S})$ 
mass in the on-shell scheme. However, in this case it is understood that 
the large coefficients are unphysical and can be removed by a suitable 
mass subtraction procedure. If a similar mechanism underlied the 
expansion for the leptonic width, the error on the bottom quark mass 
could be reduced. In the absence of any understanding of this point, 
we have adopted a more conservative error estimate than in previous works 
\cite{PP98,Hoa98,MY98,Hoa99}, mainly because of a more generous variation 
of the renormalization scale. Eq.~(\ref{msbar3}) is in agreement within 
errors, but larger than the quark mass values quoted there, 
but is about $200\,$MeV smaller than the one in \cite{PY98}. We 
argued that the result of \cite{PY98} should be corrected for 
the large 3-loop term in the relation between the pole mass and 
the $\overline{\rm MS}$ mass. 

Eq.~(\ref{msbar3}) is also in good agreement 
with $\overline{m}_b(\overline{m}_b) = 4.26 \pm 0.07\,\mbox{GeV}$ found 
in \cite{MS98}. This work uses the $B$ meson mass, a lattice calculation 
of the (properly defined) binding energy of the $B$ meson in the unquenched, 
two-flavour approximation to heavy quark 
effective theory, and a two-loop perturbative matching to the 
$\overline{\rm MS}$ scheme. To our knowledge, this is the only 
other NNLO determination of the $\overline{\rm MS}$ mass besides 
the sum rule calculations mentioned above (which, in fact, are N$^4$LO as 
far as $\overline{m}_b$ is concerned). 
However, because of the 
heavy quark limit, there are $1/m_b$ corrections, which remain to be 
estimated. Finally, the result is also in agreement with earlier, 
parametrically less accurate determinations, as for example in 
\cite{others}. 

\vspace*{0.2cm}
{\em Acknowledgements.} 
We thank G.~Buchalla, A.H.~Hoang and A.~Pineda for useful discussions 
and comments on the manuscript. 
We also thank M.~Steinhauser for providing us with the numerical code 
that produced the fixed-order moments (FO1) in Table~\ref{tab2}. This 
work was supported in part by 
the EU Fourth Framework Programme `Training
and Mobility of Researchers', Network `Quantum Chromodynamics and the Deep
Structure of Elementary Particles', contract FMRX-CT98-0194 (DG 12 - MIHT).


\begin{thebibliography}{99}

\bibitem{NSVZ}
V.A.~Novikov {\em et al.}, Phys. Rev. Lett. {\bf 38} (1977) 626 [Erratum: 
{\em ibid.} {\bf 38} (1977) 791]; {\em Phys. Rep.} {\bf 41} (1978) 1.
\bibitem{VZ87}
M.B.~Voloshin and Yu.M.~Zaitsev, Usp. Fiz. Nauk. {\bf 152} (1987) 361 
[Sov. Phys. Usp. {\bf 30}(7) (1987) 553]; \\
M.B.~Voloshin, Int. J. Mod. Phys. {\bf A10} (1995) 2865.
\bibitem{BSS98II}
M.~Beneke, A.~Signer and V.A.~Smirnov, in: Proceedings of RADCOR98, 
Barcelona, September 1998. Note that the numerical result of the version 
published in the proceedings is affected by a program error, which is 
corrected in the hep-ph archive version [hep-ph/9906476].
\bibitem{Sch98}
Y.~Schr\"oder, Phys. Lett. {\bf 447} (1999) 321; \\
M.~Peter, Phys. Rev. Lett. {\bf 78} (1997) 602.
\bibitem{BS98}
M.~Beneke and V.A.~Smirnov, Nucl. Phys. {\bf B522} (1998) 321; \\
M.~Beneke, A.~Signer and V.A.~Smirnov, Phys. Rev. Lett. {\bf 80} (1998) 
2535.
\bibitem{CM98}
A.~Czarnecki and K.~Melnikov, Phys. Rev. Lett. {\bf 80} (1998) 2531.
\bibitem{Hoa97}
A.~Hoang, Phys. Rev. {\bf D56} (1997) 7276.
\bibitem{Ben98a}
M.~Beneke, Talk given at 33rd Rencontres de Moriond: Electroweak Interactions 
and Unified Theories, Les Arcs, France, 14-21 Mar 1998 [hep-ph/9806429]. 
\bibitem{BSS99}
M.~Beneke, A.~Signer and V.A.~Smirnov, Phys. Lett. {\bf B454} (1999) 137.
\bibitem{Ben98}
M.~Beneke, Phys. Lett. {\bf B434} (1998) 115.   
\bibitem{HSSW98}
A.H.~Hoang, M.C.~Smith, T.~Stelzer and S.~Willenbrock, 
Phys. Rev. {\bf D59} (1999) 114014.
\bibitem{BB94}
M.~Beneke and V.M.~Braun, Nucl. Phys. {\bf B426} (1994) 301; \\
M.~Beneke, Phys. Lett. {\bf B344} (1995) 341; \\
I.I.~Bigi, M.A.~Shifman, N.G.~Uraltsev and A.I.~Vainshtein, Phys. Rev. 
{\bf D50} (1994) 2234. 
\bibitem{PP98}
A.A.~Penin and A.A.~Pivovarov, Phys. Lett. {\bf B435} (1998) 413.
\bibitem{Hoa98}
A.H.~Hoang, Phys. Rev. {\bf D59} (1999) 014039.
\bibitem{MY98}
K.~Melnikov and A.~Yelkhovsky, Phys. Rev. {\bf D59} (1999) 114009.
\bibitem{Hoa99}
A.H.~Hoang, [hep-ph/9905550].
\bibitem{PY98}
A.~Pineda and F.J.~Yndurain, Phys. Rev. {\bf D58} (1998) 094022; 
[hep-ph/9812371].
\bibitem{pdg98}
Particle Data Group, C.~Caso {\em et al.}, Eur. Phys. J. {\bf C3} (1998) 1.
\bibitem{cleo}
D.S.~Akerib {\em et al.} [CLEO-II Collaboration],
Phys. Rev. Lett. {\bf 67} (1991) 1692.
\bibitem{CKS97}
K.G.~Chetyrkin, J.H.~K\"uhn and  M.~Steinhauser, 
Phys. Lett. {\bf B371} (1996) 93; Nucl. Phys. {\bf B482} (1996) 213; 
Nucl. Phys. {\bf B505} (1997) 40.
\bibitem{CS99}
K.G.~Chetyrkin and M.~Steinhauser,
[hep-ph/9907509].
\bibitem{BBB}
M.~Beneke and V.M.~Braun, Phys. Lett. {\bf 348} (1995) 513; \\
P.~Ball, M.~Beneke and V.M.~Braun, Nucl. Phys. {\bf B452} (1995) 563.
\bibitem{Neu95}
M.~Neubert,
Phys. Rev. {\bf D51} (1995) 5924; \\
C.N.~Lovett-Turner and C.J.~Maxwell,
Nucl. Phys. {\bf B452} (1995) 188.
\bibitem{LV81}
H.~Leutwyler, Phys. Lett. {\bf B98} (1981) 447; \\
M.B.~Voloshin, Sov. J. Nucl. Phys. {\bf 36}(1) (1982) 143.
\bibitem{Pin97}
A.~Pineda, Nucl. Phys. {\bf B494} (1997) 213.
\bibitem{JP98}
M.~Jamin and A.~Pich,
Nucl. Phys. Proc. Suppl. {\bf 74} (1999) 300 [hep-ph/9810259];
Nucl. Phys. {\bf B507} (1997) 334.
\bibitem{HLM99}
A.H.~Hoang, Z.~Ligeti and A.V.~Manohar,
Phys. Rev. {\bf D59} (1999) 074017.
\bibitem{MS98}
V.~Gim\'{e}nez, L.~Giusti, F.~Rapuano and G.~Martinelli, 
[hep-lat/9909138].
\bibitem{others}
L.J.~Reinders, H.~Rubinstein and S.~Yazaki,
Phys. Rept. {\bf 127} (1985) 1;
S.~Narison, Phys. Lett. {\bf B341} (1994) 73.
\end{thebibliography}
\end{document}